\documentclass{elsart1p}
 \usepackage{graphics}
\usepackage{epsfig}
\usepackage{graphicx}

\usepackage{amssymb}
\begin{document}
\begin{frontmatter}
% use the corauthref command within \author for corresponding author
% footnotes;
%\title{Relativistic Chiral Theory of Nuclear Matter and QCD Constraints}{label1}}
 %\thanks[label1]{IPN Lyon, IN2P3/CNRS, Université Lyon1, France}
% \author{G. Chanfray\corauthref{M. Ericson}\thanksref{label2}}
 %\ead{email address}
% \thanks[label2]{IPN Lyon, IN2P3/CNRS, Université Lyon1 and Theory Division, Geneva}
 %\corauth[cor1]{}
 %\thanks[label3]{}
\title{Relativistic Chiral Theory of Nuclear Matter and QCD Constraints}
 \author[label1]{G. Chanfray}
 \author[label1,label2]{M. Ericson}
 \address[label1]{IPN Lyon, IN2P3/CNRS, Université Lyon1, France}
 \address[label2]{ Theory Division, CERN, Geneva, Switzerland}
%\author{}
%\address{}
%
\begin{abstract}
We present a relativistic chiral theory of nuclear matter which includes the effect of confinement.
Nuclear binding is obtained  with a chiral invariant scalar background field  associated with the radial fluctuations of the chiral condensate
Nuclear matter stability is ensured once the scalar response of the nucleon depending on the quark confinement mechanism is properly incorporated. All the parameters are fixed or constrained by  hadron phenomenology and lattice data. A good description of nuclear saturation  is reached, which includes the effect of in-medium pion loops. Asymmetry properties of nuclear matter are also well described once the full rho meson exchange 
and Fock terms are included.

\end{abstract}
\begin{keyword}
Chiral symmetry, confinement, nuclear matter
\PACS{24.85.+p 11.30.Rd 12.40.Yx 13.75.Cs 21.30.-x} 

\end{keyword}
\end{frontmatter}
%
% main text
%\section{Deadline for submission}

\noindent
{\it General context.} The aim of this talk is to discuss the possible relation between fundamental properties of low energy QCD, namely chiral symmetry and confinement, with the  rich structure of the nuclear many-body problem. More precisely with the introduction of the concept of nucleon substructure adjustement to the nuclear environment \cite{G88} we will bring some constraints from lattice QCD data into the modelling of many-body forces.
In  the relativistic mean-field approaches initiated by Walecka, the nucleons move in an attractive scalar  and in a repulsive vector background fields. This provides a very economical saturation mechanism and a spectacular well known success is the correct magnitude of the spin-orbit potential since the large vector and scalar fields contribute to it in an an additive way. Now the question of the very nature of these background fields  and their  relationship with the QCD condensates has to be elucidated. To address this problem we formulate the effective theory  in terms of the fields associated with the fluctuations of the chiral condensates in a matrix form ($W=\sigma + i\vec\tau\cdot\vec\pi$) 
by going from cartesian  to polar coordinates  {\it i.e.}, going from a linear to a non linear representation~:
$W=\sigma\, + \,i\vec\tau\cdot\vec\pi=S\,U=(f_\pi\,+\,s)\,exp\left({i\vec\tau\cdot\vec\varphi_\pi/ f_\pi}\right)$.  In ref \cite{CEG02} we made the physical assumption  to identify the chiral invariant scalar field $s=S - f_\pi$, associated with {\it radial} (in order to respect chiral constraints) fluctuations   of the condensate, with the background attractive scalar field.

In this approach the Hartree energy density of nuclear matter, including $\omega$ exchange, writes in terms of the order parameter $\bar s=\langle s \rangle$~:
$E_0/ V=\varepsilon_0=\int\,(4\,d^3 p /(2\pi)^3) \,\Theta(p_F - p)\,E^*_p(\bar s)
\,+\,V(\bar s)\,+\,g^2_\omega/ 2\, m_{\omega^2}\,\rho^2$,
where $E^*_p(\bar s)=\sqrt{p^2\,+\,M^{*2}_N(\bar s)}$ is the energy of an 
effective nucleon with the effective Dirac mass $M^*_N(\bar s)=M_N +g_S\,\bar s$. $g_S=M_N/f_\pi$ is the scalar coupling constant of the sigma model. Here two serious problems appear. The first one is  the fact that the chiral effective potential,
$V(s)$, contains an attractive tadpole diagram which  generates an attractive three-body force destroying  matter stability. The second one is related to the nucleon substructure. The nucleon mass can be expanded according to
$M_N(m^{2}_{\pi}) = 
a_{0}\,+\,a_{2}\,m^{2}_{\pi}\, +\,a_{4}\,m^{4}_{\pi}\,+\,\Sigma_{\pi}(m_{\pi}, \Lambda)+...$, where the pionic self-energy is explicitely separated out. The $a_2$ parameter is related to the non pionic 
piece of the $\pi N$ sigma term and $a_4$ to the nucleon QCD scalar susceptibility. According to the lattice data analysis $a_4$ is found to be negative, $(a_4)_{latt} \simeq- 0.5\, \mathrm{GeV}^{-3}$ \cite{LTY04}, but much smaller than in our chiral effective model,
$\left(a_4\right)_{Chiral}=-3f_{\pi} g_S/2\,m^{4}_{\sigma}\simeq -3.5\,GeV^{-3}$, where the nucleon  is seen  as a juxtaposition of three constituent quarks getting their mass  from the chiral condensate. The common origin of these two failures can be attributed to the absence  of confinement. In reality the composite nucleon should respond to the nuclear environment, {\it i.e.}, by readjusting its confined quark structure. The resulting  polarization of the nucleon is accounted for by the phenomenological introduction of the scalar nucleon response, $\kappa_{NS}$, in  the nucleon mass evolution ~: $M_N(s)=M_N\,+\,g_S\,s\,+\,\frac{1}{2}\,\kappa_{NS}\,s^2\,+\,....$.
This constitutes the only change in the expression of the energy density  but this has numerous consequences. In particular
the $a_4$ parameter is modified~: $a_4=\left(a_4\right)_{Chiral}\left(1\,-\,\frac{2}{3} C\right)$. The value of  $C\equiv(f^{2}_{\pi}/2\,M_N)\kappa_{NS}$ which reproduces the lattice data is $C\simeq 1.25$ implying a strong cancellation effect in $a_4$. Moreover  the scalar response of the nucleon induces 
an new piece in the lagrangian ${\cal L}_{s^2 NN=}-\,\kappa_{NS}\,s^2\,\bar N N/2$ which  generates a repulsive three-body force able to restore saturation.

\smallskip\noindent
{\it Applications.}  The restoration of saturation properties has been confirmed at the Hartree level \cite{CE05} with a value of the dimensionless scalar response parameter, $C$,  close to the value estimated from the lattice data. The next step has been to include pion loops on top of the Hartree mean-field calculation \cite{CE07}. We use a standard many-body (RPA) approach which includes the effect of short-range correlation ($g'$ parameters). Non relativistically, the main ingredient is the full polarization propagator $\Pi_L$ (which also includes the $\Delta-h$ excitations)  in the longitudinal spin-isospin channel. To be complete we also add the transverse spin-isospin channel ($\Pi_T$) and the associated rho meson exchange. The calculation of $E_{loop}=E\,-\,E_0$ is done using the well-known charging formula with residual interaction $V_L=\pi+g',\, V_T=\rho+g'$~:
$$
\frac{E_{loop}}{V}=3\,V\,\int_{-\infty}^{+\infty} 
{i\,d\omega\over 2\pi)}\int{d{\bf q}\over (2\pi)^3}\,\int_0^1{d\lambda\over\lambda}
\left( V_L(\omega, {\bf q}) \Pi_L(\omega, {\bf q}; \lambda)\,+\,2\,
V_T(\omega, {\bf q}) \Pi_T(\omega, {\bf q}; \lambda)\right)~.\label{ELOOP}
$$
The parameters associated with spin-isospin physics  are fixed by nuclear phenomenology  and  the calculation has no real free parameters apart for a fine tuning for $C$ (around the lattice estimate) and for  $g_\omega$  (around the VDM value). The result of the calculation is shown on fig. 1. We stress   the relatively modest value of the correlation energy ($-8\, MeV$ and $-7 \,MeV$ for the L and T channels), much smaller that what is obtained from iterated pion exchange (planar diagramm) in  in-medium chiral perturbation theory. This effect is mainly due to the strong sceening of pion exchange from short-range correlations.

We have also performed  a full relativistic Hartree-Fock (RHF) calculation with the notable inclusion of the rho meson exchange which is important to also reproduce the asymmetry properties of nuclear matter \cite{MC08}. Again in an almost parameter free calculation, saturation properties of nuclear matter can be reproduced with 
$g_S=10$, $m_\sigma=800\, MeV$ (lattice), $g_\rho=2.6$ (VDM), $g_\omega=6.4$ (close to the VDM value $3\,g_\rho$) and $C=1.33$ (close to the lattice value). For the tensor coupling we a first take the VDM value $\kappa_\rho=3.7$.  The corresponding  asymmetry energy, $a_S$, is shown on the right panel of fig 1. It is important to stress that the rho Hartree contribution ($7\,MeV$) is not sufficient  when keeping the VDM value for the vector coupling constant $g_\rho$. The Fock term through its tensor contribution is necessary to reach the range of accepted value of $a_S$ around $30\, MeV$. Increasing $\kappa_\rho$ (strong rho scenario) to $\kappa_\rho=5$ allows a particularly good reproduction of both symmetric and asymmetric nuclear matter. The model also predicts a neutron mass larger than the proton mass with increasing neutron richness in agreement with ab-initio BHF calculations \cite{SBK05}.
This  approach gives very encouraging results and raises  question of how confinement can  generate such a large and positive scalar response of the nucleon, $\kappa_{NS}$. This is presumably linked to a delicate balance of (partial) chiral symmetry restoration and confinement mechanism inside the nucleon.
\noindent
\begin{figure}
  \begin{tabular}{cc}
  \begin{minipage}{.50\linewidth}
    \includegraphics[scale=0.3,angle=270]{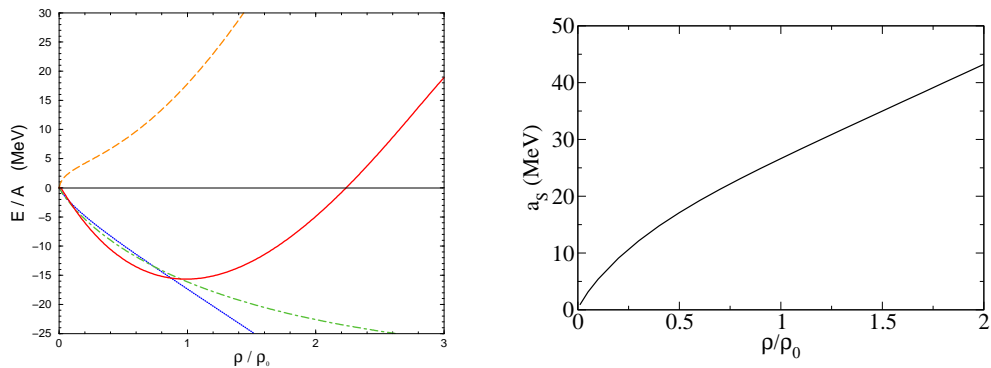}
  \end{minipage}
  &
  \begin{minipage}{.50\linewidth}
%    \centering\epsfig{figure=asy-1.eps,width=\linewidth}
\includegraphics[scale=0.25]{asy-1.eps}
   \end{minipage}
   \end{tabular}  
\caption{Left panel: Binding energy of nuclear matter with $g_\omega=8$, $m_\sigma =850$ MeV and $C=0.985$ 
with the Fock  and correlation energies on top of $\sigma$ and
$\omega$ exchange. Full line:  full result. Dotted line: 
without Fock and  correlation energies. Dot-dashed line:  Fock terms. Decreasing dotted line (always negative):
correlation energy \cite{CE07}.
Right panel: Asymmetry energy versus density in the RHF approach  \cite{MC08}.}
\end{figure}

\label{}
% \appendix
%
% \section{}
% \label{}
%

%
\end{document}